\documentclass[12pt]{iopart}
\usepackage{iopams}
\usepackage{dsfont}
\usepackage{graphicx,verbatim}
\usepackage{color}

\begin{document}

\title[]{Non-adiabatic quantum pumping by a randomly moving quantum dot}

\author{Stanislav Derevyanko}

\address{Department of Physics of Complex Systems, Weizmann Institute of Science, Rehovot 76100, Israel}
\ead{stanislav.derevyanko@weizmann.ac.il}

\author{Daniel Waltner}

\address{Faculty of Physics, University of Duisburg-Essen, 47048 Duisburg, Germany}
\ead{daniel.waltner@uni-due.de}

\begin{abstract}
We look at the  time dependent fluctuations of the electrical charge in an open 1D quantum system represented by a quantum dot experiencing random lateral motion. In essentially
non-adiabatic settings we study both diffusive and ballistic (Levy) regimes of the barrier motion where the electric current as well as the net pumped electric charge experience random
fluctuations over the static background. We show that in the large-time limit, $t \to \infty$, the wavefunction is naturally separated into the Berry-phase component (resulting from the singular
part of the wave amplitude in the co-moving frame) and the non-adiabatic correction (arising from fast oscillating, slow decaying tails of the same amplitude). Based on this separation we report two key results: Firstly, the disorder averaged wave function and current are asymptotically
mainly determined by the same Berry phase contribution that applies in the case of adiabatic motion. Secondly, after a short transition period the pumped electric charge exhibits fluctuations that grow much faster than predicted by the adiabatic theory.
We also derive the exact expressions for the average propagator (in the co-moving basis representation) for the diffusive and ballistic types of motion considered.
\end{abstract}

\submitto{\JPA}
\maketitle

\section{Introduction}
Since the pioneering work of Thouless \cite{t83} quantum pumping, i.e. transporting charge via a time-dependent potential or potential barrier (rather than applying voltage bias) has attracted great
attention in the theory of mesoscopic devices, chiefly with the goal to mediate transport in Josephson junctions \cite{gigsj01,sggij03,rn04} and quantum dots \cite{btp94,b98,aegs00,mb02,k02,k04,jcsf08,cgk09}. The simplest theoretical framework for the quantum pumping phenomenon is provided by a 1D
quantum scattering at a time-dependent potential. The overwhelming majority of the existing results on time dependent quantum scattering were obtained assuming that either the time dependence of the
potential barrier is due to periodic variations of its height and/or position, \cite{mb02,k02,k04,jcsf08,phs91,hdr00,gh04,la05,bipamdd12} or the changes are adiabatic so that at each
time moment one can use a static scattering matrix \cite{btp94,b98,aegs00} to calculate the net transported charge. Some notable exceptions from these frameworks were considered in \cite{as07} where
non-adiabatic pumping was considered in a discrete tight-binding model with harmonic on-site potential and \cite{cks05} where corrections to adiabaticity were obtained using a Kubo formula approach (see also Ref. \cite{cgk09} for the non-adiabatic spin pumping).
Some of the numerical results reported in \cite{jcsf08} also concern the non-adiabatical pumping but with AC variations of both height and position
of the potential. Additionally, in Ref.\ \cite{ws14} the impact of a randomly varying bond length in a quantum graph on the current was considered.

In this paper we want to study quantum pumping in new settings where it results from non-stationary scattering at a narrow barrier (modeling quantum dot) experiencing a random walk in time, while preserving its shape. We consider a continuous time random walk (CTRW) where the scatterer can move either diffusively, via Brownian motion (BM), corresponding to delta correlated Gaussian velocity or with a constant velocity between the consecutive turning events with the ``fat-tailed'' distribution between these events (Levy walk)  \cite{zk93,KS}. The latter case generally corresponds to superdiffusion. Both cases (especially the BM) provide bona fide examples of a non-adiabatic change of the parameters. Our goal here will be twofold: first, we will show that if one is interested in the transported charge then one can get a non-zero fluctuating value via lateral translation
only (since periodicity is lost). This net charge should have zero average for unbiased CTRW and here we study how its variance grows with time.

Our second goal is to show that in the non-adiabatic limit of CTRW an arbitrary initial state does evolve into the combination of a Berry contribution that follows the change
of the scatterer position up to the same geometric phase factor \cite{b84} (see also \cite{aegs00}) and the non-adiabatic correction that leads to fast fluctuations of the probability current. The amazing
result is however that when one considers the \textit{disorder-averaged} values of the wave function and the probability current then asymptotically these seem to be solely determined by the
Berry phase contribution. We call this result non-adiabatic Berry phase (BP) regime. In this asymptotic regime the average electric current is almost uniform and the growth of
the fluctuations of the pumped charge is determined by the non-adiabatic fluctuations only. We show numerically that these fluctuations grow faster than predicted by the commonly
used adiabatic "snow-plow" formula. As a by-product of our analysis we also obtain closed form expressions for the disorder-averaged quantum propagator (in the co-moving frame representation) for
both models of the CTRW. Finally we note that although the role of the BP in the quantum pumping and quantum ratchets has been studied before \cite{sn07,sn07-01,yssh12,yh13} (largely by considering some form of a quantum master equation) our paper presents the first results (to the best of our knowledge) on the \textit{strongly non-adiabatic} stochastic quantum pumping in mesoscopic systems.

\section{Problem statement and the model}
We are considering a 1D quantum dot placed between two leads at zero temperature at thermodynamic equilibrium. No external gate voltage is applied so both leads have the same value of chemical potential corresponding to the Fermi level: $\mu_-=\mu_+=E_F$. Let us assume that the quantum dot experiences only lateral motion without any shape change of the potential. This can be induced e.g. by external AC irradiation of the scatterer.
The situation can be modeled by the 1D non-stationary Schr\"odinger equation with a given potential barrier $V(x)$ experiencing a lateral shift in time:
\begin{equation}
i\,\hbar \,\frac{\partial \psi}{\partial t}= -\frac{\hbar^2}{2m}\,\psi''+V(x-\gamma(t))\,\psi
\label{Schrodinger}
\end{equation}
where $\gamma(t)$ is a time dependent position of the scatterer and we assume that $\gamma(0)=0$ always i.e. at $t=0$ the system is fully characterized by the eigenstates of the static
problem. We will assume here that the barrier is narrow compared to the typical wave-length of the electron wave (which is the Fermi wave-length, $2\pi/k_F$). Without
loss of generality we will consider a delta-shaped repulsive scatterer, although in Section \ref{sec:prop} we also present some generic results valid for an arbitrary shape of the barrier.  As for
the position shift $\gamma(t)$ we will assume that it experiences a CTRW of a particular kind. Specifically we assume that the barrier is moving with a piece-wise constant velocity and from time to
time experiences a ``jolt'' which instantaneously changes the velocity to a new value after which the system continues to move with the new velocity until the next jolt, etc. We will assume that
jolts represent events connected to a random process with independent stationary increments. Two different models are considered. In the first (Poissonian) model the time between the jolts (i.e. waiting time) is
exponentially distributed and its average, $\tau$, plays the role of the velocity correlation time. As for the velocities we will assume that at each jolt the new velocity
is chosen from a zero mean Gaussian distribution with the variance $\sigma_v^2=2D/\tau$, with some positive constant $D$. The second case is the two state velocity model \cite{zk93} (or Levy walk in the terminology
of \cite{KS}) where the inter-event time has a heavy tailed distribution (i.e. decays as $\tau^\alpha/t^{1+\alpha}$, $0<\alpha<1$ as $t\to \infty$) while the velocity at each event changes direction and can only take two values $\pm v$. For the range of the exponents $\alpha$ given above all the momenta for the waiting time distribution starting from the first diverge and this process does not have a well defined characteristic scale \cite{KS}.

The reason why these two particular models are chosen is that for the former model, in the limit $\tau \to 0$ (which will be studied here) one obtains the continuous white noise model for the
velocity:
 \begin{equation}
\label{Brownian}
\gamma(t)=\intop_0^t\,\xi(t')\,dt', \quad \overline{\xi(t)} =0, \quad  \overline{\xi(t) \,\xi(t')} =2 D\,\delta(t-t').
\end{equation}
This corresponds to the unbiased \textit{diffusive} BM with the familiar Einstein's relation for the time growth of the root mean square displacement (RMS): $\sqrt{\overline{\gamma^2(t)}}=\sqrt{2D t}$.
On the other hand it is known that the Levy walk (in the specified range of the exponents $\alpha$) is \textit{ballistic} with the spread velocity $V=(1-\alpha)^{1/2} v$. In all our
numerical simulations we have chosen the so-called Levy-Smirnov distribution for the inter-event time, $p(t)$, which corresponds to $\alpha=1/2$ \cite{KS}. Let us stress that we will assume that in
both models the parameters of the CTRW are essentially \textit{non-adiabatic} and the scatterer position experiences rapid random fluctuations.

The solution $\psi(x,t)$ at an arbitrary moment of time can be presented as an expansion over Galilei shifted static eigenstates in the co-moving frame (see e.g. \cite{phs91}):
\begin{equation}
\psi(x,t)= \int dk \, c(k,t) \,\psi_k(x-\gamma(t))\, e^{-\frac{i E(k)}{\hbar} \,t}, \quad E(k)=\frac{\hbar^2 k^2}{2 m}.
\label{expansion}
\end{equation}
The above equation defines a new time dependent shifted basis $|k, \gamma(t)\rangle$ obtained from the eigenvectors of the static problem $|k, 0 \rangle$ via unitary transformation:
\begin{equation}
\label{rotation}
|k,\gamma(t)\rangle =\hat{U}(t)| k, 0 \rangle, \quad \hat{U}(t) \equiv \exp\left[ -\frac{i}{\hbar}\,\hat{p} \gamma(t)- \frac{i}{\hbar} \hat{H_0} t \right],
\end{equation}
where $\hat H_0$ is the static part of the Hamiltonian in (\ref{Schrodinger}) and $\hat{p}$ is the 1D momentum operator. In what follows we will mostly omit the $\gamma$-dependence for brevity and use shorthand notation $|k \rangle$ for the co-moving basis. Clearly in the coordinate representation one has $\langle x| k \rangle \equiv \psi_k(x-\gamma(t)) \exp[-i E(k) t/\hbar]$.

Of particular importance to the problem of the non-stationary quantum transport are the so-called \textit{scattering states} $\chi_{k_0}$ \cite{k04,lw99}. These are defined as the solutions of time dependent equation (\ref{Schrodinger}) that at $t=0$ coincide with the static eigenfunctions $\psi_{k_0}(x)$. In the co-moving frame representation (\ref{expansion}) the scattering states correspond to the initial condition $c(k,0)=\delta(k-k_0)$. Another interpretation of a scattering state $\chi_{k_0}$ is that when projected on the moving basis $|k\rangle$ it provides the expression for quantum propagator in this representation (see Section \ref{sec:prop}). Therefore these states can also be used to study random scattering of narrow wave-packets in the $k$-space. When the potential experiences purely periodic motion starting from $t=-\infty$ the scattering states can be obtained in the closed form via the Floquet formalism \cite{k02,k04}. When the time-dependent potential is small one can also obtain
these states
perturbatively \cite{lw99}. In general however these solutions must be obtained either numerically or by making some additional assumptions e.g. considering the asymptotic limit $t \to +\infty$ which is the method we adopt in Section \ref{sec:Berry}. An important limiting case is that of the \textit{adiabatically slow motion} of the barrier. When the change of position is adiabatic then at time $t$ the scattering state is given by the corresponding moving eigenstate $|k_0 \rangle$ up to a BP factor \cite{b84,ms08}
\begin{equation}
|\chi_{k_0}(t) \rangle =e^{i\, \phi_{k_0}(t)} |k_0 \rangle.
\label{Berry}
\end{equation}
So our first goal in this paper will be to see what happens to the amplitudes $c(k,t)$ of the scattering state $\chi_{k_0}(x,t)$ when the changes of the potential are fast and random as assumed in our model. As we shall
see, in this case the complex amplitude $c(k,t)$ is characterized by sharp delta-peaks at $k=\pm k_0$ while at large times the wave function in the coordinate representation, $\chi_{k_0}(x,t)$ naturally
separates into the local Berry phase contribution (\ref{Berry}) and nonadiabatic corrections arising due to the slow (power law) decay of the amplitude $c(k,t)$ at $|k| \to \infty$. For the Levy
model the situation is more complicated since the system spends relatively large times moving with constant velocity $\pm v_0$ which lead to additional Doppler shifted peaks in $|c(k,t)|$.

We will also be interested in the properties of random quantum pumping in such a system and more specifically in the pumped charge. Traditionally when considering adiabatically slow
periodic changes of the parameters of the scattering channel (i.e. scattering matrix) the pumped charge is calculated via the well known formula of B\"{u}ttiker, Thomas and Pr\^etre (BTP) \cite{btp94}.
For a single channel system where only lateral shifts of the scatterer are considered (i.e. our situation) this formula can be considerably simplified giving for the infinitesimally small pumped
charge \cite{aegs00,jcsf08}:
\begin{equation}
\Delta Q=-\frac{e \,k_F}{\pi} |r_{k_F}|^2 \, \Delta \gamma,
\label{Buttiker}
\end{equation}
where $r_{k_F}$ is the static value of the reflection coefficient at Fermi level and $\Delta \gamma$ is the infinitesimally small change of the scatterer position. Eq.(\ref{Buttiker}) reflects the so-called
``snow-plow'' contribution to the charge \cite{aegs00}.

Since our paper concerns with the non-adiabatic and non-periodic quantum pumping one will need to simulate the electric current passing through the leads directly, without simplifying assumptions.
Using a standard procedure in which different scattering states serve as a basis for expanding the time and space dependent field operator $\hat{\Psi}(x,t)$ in terms of creation and annihilation
operators of the electron in the right and left lead and performing quantum mechanical averaging at zero temperature one arrives at the following expression for the charge pumped through the
left ``-" or right ``+" lead \cite{lw99,da92} determined at the positions $\pm L/2$:
\begin{equation}
\label{charge}
\fl Q_\pm(t) =  \intop^t_0 I_{\pm}(t') \, dt', \quad I_\pm(t)= -e \intop_{-k_F}^{k_F} J_{k_0}(\pm L/2,t) \, dk_0
\end{equation}
where $e>0$ is the elementary charge and
\begin{equation}
\label{current}
J_{k_0}(x,t)=\frac{\hbar}{2m\,i}\left[ \chi_{k_0}^*(x,t) \,\frac{\partial \chi_{k_0}(x,t)}{\partial x} -  \chi_{k_0}(x,t) \,\frac{\partial \chi_{k_0}^*(x,t)}{\partial x} \right]
\end{equation}
is the probability current (flux) associated with the particular scattering state $\chi_{k_0}$. The suitable choice of the boundaries $\pm L/2$ will be to place them outside the RMS displacement of the random potential. The physical meaning of Eq.(\ref{charge}) is transparent: the electric current passing through right (left) lead at time $t$ is proportional to the total probability flux of all right $k_0>0$ and left $k_0<0$ moving electrons under the 1D ``Fermi surface'' $[-k_F,k_F]$. For the static case $\gamma \equiv 0$ the scattering states are just phase rotated eigenstates $|\chi_{k_0}\rangle = e^{-i E(k_0) t/\hbar}|k_0 \rangle$ so that the left and right moving currents are proportional to the total transmission for the right and left movers (the celebrated Landauer-B\"uttiker formula \cite{da92,Landau,Buttiker}). In the absence of gate voltage, for any symmetric potential, the time reversal symmetry ensures that these two contributions exactly cancel each other and the total current 
in each lead is
exactly zero (see e.g. \cite{da92}). Introducing time dependent motion of the quantum dot
generally breaks time reversal symmetry, however for adiabatically slow periodic pumping the net charge accumulated over one period of oscillations is still zero \cite{jcsf08}. In our case of random non-adiabatic pumping the total charge $Q(t)$ represents a random process which is expected to have zero mean and time dependent RMS $\sigma_Q(t)$ which is the quantity we are interested in here.

Since the scattering states form the basis of the theoretical treatment of the non-adiabatic quantum pumping in the next few sections we will study in detail the statistical and asymptotic properties
of these states at arbitrary wavelength $\lambda=2\pi/k_0$ of the incoming wave. The results of the following three chapters are quite generic and can be applied beyond the context of quantum
transport in such areas as non-stationary scattering of  wavepackets \cite{bipamdd12} or quantum graphs with fluctuating bond lengths \cite{ws14}.

\section{The dynamics of scattering states}
\label{sec:ODEs}
Since each scattering state is a solution of the time dependent Schr\"{o}dinger equation (\ref{Schrodinger}) it is natural to expand it in the co-moving basis (\ref{expansion}). Let us now derive the continuous time evolution equations for the amplitudes in the co-moving frame $c(k,t)$. Inserting expansion (\ref{expansion}) into the Schr\"odinger equation
(\ref{Schrodinger}) and projecting it onto the corresponding moving eigenstate one obtains the following exact equation \cite{phs91}:
\begin{equation}
\dot{c}(k,t)=\dot{\gamma}(t)\int  A(k,k';t) \,c(k',t) \, dk', \quad c(k,0)=\delta(k-k_0).
\label{ODEs}
\end{equation}
Here the time-dependent kernel $A(k,k';t)$ is given by
\begin{equation}
\fl A(k,k';t)=  e^{\frac{i}{\hbar} \, (E(k)-E(k'))\,t} \int \psi^*_k(x)\,\frac{\partial }{\partial x} \, \psi_{k'}(x) \, dx = \langle k,0 | e^{i \hat{H}_0 t/\hbar} (i \hat{p}/\hbar)  e^{-i \hat{H}_0
t/\hbar} |k',0 \rangle.
\label{A-matrix}
\end{equation}
Note that this kernel depends only on the static Hamiltonian. Also it is interesting that Eq.(\ref{ODEs}) with the kernel (\ref{A-matrix}) looks like the evolution of the wave-function in the interaction (Dirac) representation (see e.g. \cite{M}) with the role of perturbative interaction played by the operator $\dot \gamma(t) i \hat{p}/\hbar^2$.

We are using the model of a delta-barrier $V(x)=(\hbar^2/m)\,\Omega \,\delta(x)$ with a continuous twice degenerate spectrum. Since the barrier is symmetric the static eigenfunctions corresponding to the left traveling waves $k<0$ are obtained from the right-traveling ones by the coordinate flip: $\psi_{-k}(x)=\psi_k(-x)$. The scattering is completely described by the unitary $S-$matrix:
\begin{equation}
\label{S-matrix}
 S = \left(
                      \begin{array}{cc}
                        r_k & t'_k \\
                        t_k & r'_k \\
                      \end{array}
                    \right), \quad r_k=\frac{\Omega}{i\,k-\Omega}, \quad t_k=\frac{i\,k}{i\,k-\Omega}, \quad t'_k=t_k, \quad r'_k=r_k
\end{equation}
and the left/right traveling normalized eigenfunctions are given by
\[
\psi_k(x)=(2\pi)^{-1/2}\,\left\{t_{|k|}\, e^{i k x} \theta[\pm x] + \left(e^{ikx} +r_{|k|} \,e^{-i kx}\right) \theta[\mp x] \right\},
\]
where the signs of the arguments of the Heaviside function $\theta [x]$ are chosen according to the sign of $k$ (reflecting the mirror symmetry).
From the above one readily obtains an expression for the kernel $A(k,k';0)$:
\begin{equation}
\fl A(k,k';0)=\left\{\begin{array}{cc}ik \, |t_k|^2\,\delta(k'-k)-\frac{k'(t_k^*t_{k'}-1-r_k^*r_{k'})}{2\pi(k'-k)}+\frac{k'(r_k^*+r_{k'})}{2\pi(k'+k)},&k>0,\,k'>0
\\ik \, |t_k|^2 \,\delta(k'-k)+\frac{k'(t_kt_{k'}^*-1-r_kr_{k'}^*)}{2\pi(k'-k)}-\frac{k'(r_k+r_{k'}^*)}{2\pi(k'+k)},&k<0,\,k'<0
\\-\frac{ik}{2}\,\left(t_{k}r_{k}^*-t_k^*r_{k}\right)\delta(k'+k)+\frac{k'(r_k^*t_{k'}^*-r_{k'}^*t_{k}^*)}{2\pi(k'+k)}+\frac{k'(t_{k'}^*-t_{k}^*)}{2\pi(k'-k)},&k>0,\,k'<0
\\-\frac{ik}{2} \,\left(t_{k}^*r_{k}-t_k r_{k}^*\right)\delta(k'+k)+\frac{k'(r_{k'}t_{k}-r_{k}t_{k'})}{2\pi(k'+k)}+\frac{k'(t_{k}-t_{k'})}{2\pi(k'-k)},&k<0,\,k'>0
\end{array}\right. .
\label{Akk}
\end{equation}
Generally for any symmetric potential it follows that the kernel is asymmetric with respect to inversion: $A(-k,-k';t)=-A(k,k';t)$. This means that if we know the solution of Eq.(\ref{ODEs}), say $c(k,t)$, then $c(-k,t)$ is the solution of the time-reversed equation Eq.(\ref{ODEs}) with $\dot{\gamma}(t) \to -\dot{\gamma}(t)$. Also since the Hamiltonian $\hat{H_0}$ generally does not commute with the momentum $\hat{p}$ the matrix $\hat{A}(t)$ is always non-diagonal.

Another technical note is that if $\gamma(t)$ is the Brownian motion, (\ref{Brownian}), then Eq.\ (\ref{ODEs}) represents a stochastic integro-differential equation with \textit{multiplicative
white noise}. This means that one can understand it either in Ito or in Stratonovich sense \cite{G}. Here we will opt for the latter, which actually implies that the two consecutive jolts separated
by infinitesimally small average time moments are correlated - which seemingly contradicts the independence assumption postulated above in our model. We argue however that this discrepancy (expressed in the additional spurious drift \cite{G}) is not crucial
for the observed physical effects and is simply an artifact of the chosen model. Indeed the latter can be always amended to include correlated velocity increment to yield the Stratonovich picture in
the limit $\tau \to 0$.

\section{The asymptotic Berry phase}
\label{sec:Berry}

In this section we shall mostly concentrate on the asymptotic behaviour of the solution $c(k,t)$ of (\ref{ODEs}) for a given realization of the CTRW $\gamma(t)$. The ensemble averaged scattering
states are given in the next section. Many of the results of this section are valid for arbitrary continuous translations of the barrier and may have applications beyond the scope of
quantum transport.

For an arbitrary barrier shape it is not possible to obtain the solution of (\ref{ODEs}) in a closed form. However an important simplification comes from considering the structure of the
time-dependent kernel $A(k,k';t)$ in Eqs.\ (\ref{A-matrix},\ref{Akk}). Indeed one can see that the first two terms in $A(k,k';0)$ always contain singularities at $k=\pm k'$ (the delta-like singularity and
a simple pole) so that the main contribution of these terms in the r.h.s. of Eq.(\ref{ODEs}) comes from the vicinity of these poles. In particular if one starts with the pure (right-moving)
eigenstate $|k_0\rangle$, than instantly the opposite, mirror-reflected eigenstate $|-k_0\rangle$ is excited. These states continue to move in space, dragged by the barrier, so that at any given time
moment the amplitude in the co-moving frame $c(k,t)$ will contain the two peaks of variable height at $k=\pm k_0$. But in the vicinity of these peaks one can neglect the oscillating term
$\exp[i\,(E(k)-E(k'))t/\hbar]$ in the definition of the kernel $\hat{A}$ (\ref{A-matrix}) i.e. assume that $\hat{A}(t) \approx \hat{A}(0)$. This approximation fails to describe the tails of the
amplitude $c(k,t)$ and, as we shall see below in the limit of large time these tails are responsible for the break-up of the adiabatic BP ansatz.

For the static kernel $\hat{A}(0)$ the solution of Eq.(\ref{ODEs}) is expressed via the usual (not time ordered) exponential propagator $\exp[\gamma(t) \hat{A}(0)]$. As mentioned before the operator $\hat{A}(0)=i \hat{p}/\hbar$ is not diagonal in the $|k,0\rangle$-basis since it generally does not commute with the Hamiltonian. This difficulty can of course be overcome by changing into the continuous eigensystem $|\lambda \rangle$ of the momentum operator which is just a set of normalized exponentials $\exp[i \lambda x]$ . In this basis we can easily calculate the propagator \textit{analytically} and then return into the original  $k$ representation to get the solution of Eq.(\ref{ODEs}).

It is straightforward to show that if initially the system is in the eigenstate $|k_0 \rangle$, then the amplitude at the time moment $t$ is given by:
\begin{equation}
c(k,t)= \langle k | t \rangle  = \int d\lambda \, e^{i\,\lambda \, \gamma(t)} \, \langle k | \lambda \rangle \, \langle \lambda | k_0 \rangle,
\label{c-answer}
\end{equation}
where real line singularities (if any) should be treated in the sense of Cauchy principal value.

Next using the static wave-functions of our delta-barrier in the coordinate domain, $\psi_k(x)$ one obtains:
\begin{eqnarray*}
\fl \langle k| \lambda \rangle & = & \int dx \, \psi_k^{*}(x) \, \frac{1}{\sqrt{2\pi}}\,e^{i\,\lambda x} =\frac{i\,|k|\,r^*_{|k|}}{\pi(\lambda^2-k^2)} +\frac{1}{2}(1+t_{|k|}^*)\delta(\lambda-k)
 +\frac{1}{2} \,r_{|k|}^* \, \delta(\lambda+k) \\
 \fl \langle \lambda| k \rangle  &=& \langle k| \lambda \rangle^*.
\end{eqnarray*}
Using the expression above and performing Fourier transform with respect to $\lambda$ (Eq.(\ref{c-answer})) we finally arrive at the following answer:
\begin{equation}
\label{c-answer1}
\fl
\begin{array}{ll}
c(k,t) &=f(k_0,t)\,\delta(k-k_0) + g(-k_0,t)\,\delta(k+k_0)\\
& + \frac{1}{2\pi \, (k^2-k_0^2)}\, \left \{2 \,\mathrm{Sign}[k] \, r^*_{|k|} \, r_{k_0}\left[k\,\sin(k_0 |\gamma(t)|) - k_0\,\sin(k |\gamma(t)|) \right]  \right.\\
& - i\,k_0 r_{k_0}\left[(1+t^*_{|k|}+r^*_{|k|})\,\cos(k \gamma(t)) + i\,(1+t^*_{|k|}-r^*_{|k|})\,\sin (k \gamma(t)) \right] \\
& \left. - i\,|k| r_{|k|}\left[(1+t_{k_0}+r_{k_0})\,\cos(k_0 \gamma(t)) + i\,(1+t_{k_0}-r_{k_0})\,\sin (k_0 \gamma(t)) \right] \right\}, \\
\end{array}
\end{equation}
where in order to obtain the expression above we have made use of the identity:
\begin{eqnarray*}
 &\mbox{p.v.} & \int_{-\infty}^\infty \frac{e^{i\lambda \gamma}}{(\lambda^2-k^2)(\lambda^2-k_0^2)}\, d\lambda \\
&=&  \frac{\pi}{k\,k_0} \, \frac{k\sin(k_0|\gamma|)-k_0\sin(k|\gamma|)}{k^2-k_0^2} + \frac{\pi^2}{2k_0^2} \, \left(\delta(k-k_0)+\delta(k+k_0)\right)\,\cos{k_0\,\gamma(t)},
\end{eqnarray*}
where the regular part of the expression above can be easily obtained via residues and the delta-part (corresponding to coalescing poles) can be calculated by integrating the l.h.s. over two infinitesimal semi-infinite stripes $|k \pm k_0| < \epsilon$, interchanging the order of $k$ and $\lambda$ integration and letting $\epsilon \to 0$.

The amplitudes $f(k_0,t)$ and $g(-k_0,t)$ have simple closed-form expressions (for a delta potential):
\begin{equation}
\label{Weyl}
\fl f(k_0,t)=\cos(k_0\,\gamma(t))+i\,\frac{k_0^2}{k_0^2+\Omega^2}\,\sin(k_0 \gamma(t)), \quad g(-k_0,t)=-\frac{k_0 \Omega}{k_0^2+\Omega^2}\,\sin(k_0 \,\gamma(t)).
\end{equation}
and correspond to unnormalized Weyl differentials \cite{ms08,M} in the moving basis representation of the initial state and its mirror counterpart.

One can see that the result (\ref{c-answer1}),(\ref{Weyl}) is quite far from the adiabatic BP formula (\ref{Berry}) as the states different from $|k_0\rangle$ are excited and the amplitude of the state itself, $f(k_0,t)$, (which has a natural meaning of survival amplitude) is not a pure phase factor (unless the scatterer is very weak, $\Omega \ll k_0$). But as we shall see below, when one considers the wave function in the coordinate domain the result (\ref{Berry}) is largely recovered.

At this stage it is pertinent to see how well the results above describe the genuine behaviour of the amplitude $c(k,t)$. To this end we now turn to numerical simulations performed within the framework of the
delta-scattering potential model (\ref{S-matrix})-(\ref{Akk}). We have solved Eq.(\ref{ODEs}) assuming that $k$-space is discretized in the range $[-3k_0,3k_0]$ using 4096 discrete points with the
grid size, $\Delta k$, imposing an effective box quantization with the box size $L'=2\pi/\Delta k$. The spatial coordinates were measured in units of
$1/k_0$ while time was measured in units of the wave-period $T=2\pi \hbar/E(k_0)$. The amplitudes themselves were also rescaled:  $c(k,t) \to c(k,t)\,\Delta k$ so that the initial condition $c(k,0)=\delta(k-k_0)$
now corresponds to a single discrete peak $\delta_{k\, k_0}$. In all our simulations we have chosen the values of the parameters that correspond
to the non-perturbative regime and the scattering was ``strong'': $\Omega=0.5 k_0$. As for the specifics of the CTRW we used (in the selected units) the values $D=0.5$ for the BM while for the Levy walk
the constant velocity $v_0$ was chosen in such a way that for the corresponding wave-vector $\tilde k= m v_0/\hbar$ one has $\tilde k =0.3 k_0$. For the Levy walk the velocity changed sign after random time
moments sampled from the Levy-Smirnov probability density function (PDF) $P(t)=\tau^{1/2}/(2\sqrt{\pi} t^{3/2})\,\exp(-\tau/4t)$ with $\tau=4\times 10^{-3}$ (which is the same as the smallest time grid in the diffusive case).

\begin{figure}[h!]
\centering\includegraphics[width=14cm]{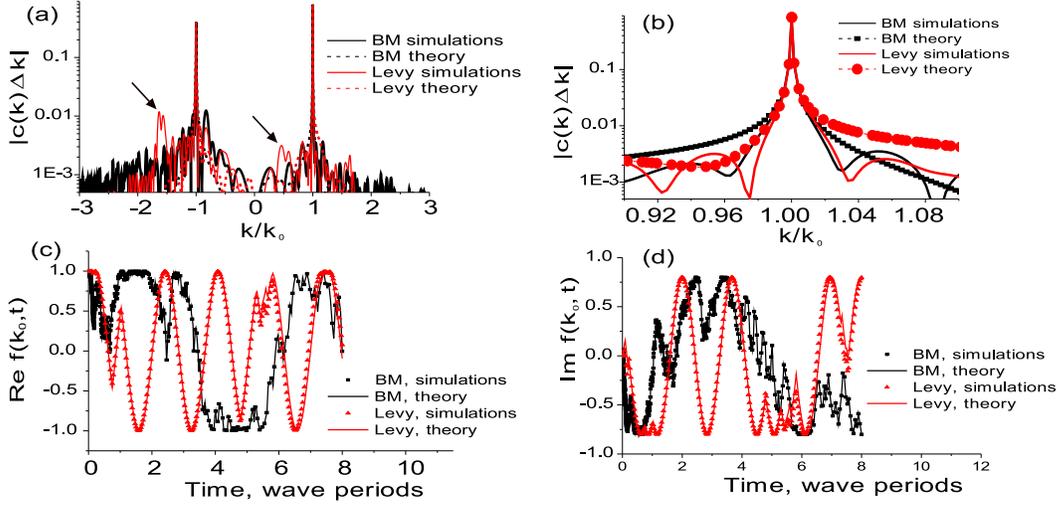}
\caption{\label{fig:compare_new} The comparison of the modulus of the numerical amplitude $|c(k,t)|\Delta k$ after $t=8$ wave periods, (a),(b) and the survival amplitude $f(k_0,t)$, (c),(d), with Eqs.(\ref{c-answer1})-(\ref{Weyl}).}
\end{figure}

The results for the amplitudes $c(k,t)$ and $f(k_0,t)$ are given in Fig.\ref{fig:compare_new} for two separate realizations of the Brownian motion and Levy walk. The first result of the comparison is that Eq.(\ref{c-answer1}) indeed provides a very good approximation in the vicinity of the peaks $k=\pm k_0$. To demonstrate it in Fig.\ref{fig:compare_new}(b) we have enlarged the area around the main peak $k=k_0$. Both real and imaginary part of $f(k_0,t)$ are in excellent agreement with the predictions of (\ref{Weyl}) as seen from  Fig.\ref{fig:compare_new}(c),(d).

However outside the peaks the difference $\tilde c(k,t)$ between the genuine amplitude $c(k,t)$ and the approximation $c_0(k,t)$ provided
by Eq.(\ref{c-answer1}) is obvious. One can see that $\tilde c(k,t)$ is a regular, fast oscillating function with slow decaying power-law
tails. Another feature concerns the difference between Brownian and Levy motion. First of all the Levy case is characterized by additional Doppler shifted peaks located at $k=\pm|k_0 \pm 2\tilde k|$.
These peaks characterize the motion with constant velocity $\pm v_0$ and are discussed in more detail in section \ref{sec:prop} (see also Ref.\cite{jcsf08}). The most significant of these peaks are
denoted by arrows in Fig.\ref{fig:compare_new}(a). Note that since for the BM case the average instantaneous velocity (or $\tilde k$) is zero these peaks are absent. Also since in the Levy case the system spends a
relatively large amount of time locked into a constant velocity motion a large part of the corresponding graphs \ref{fig:compare_new}(c),(d) displays just sinusoidal oscillations as follows
from (\ref{Weyl}).
\vspace*{3mm}
\begin{figure}[h!]
\centering\includegraphics[width=10cm]{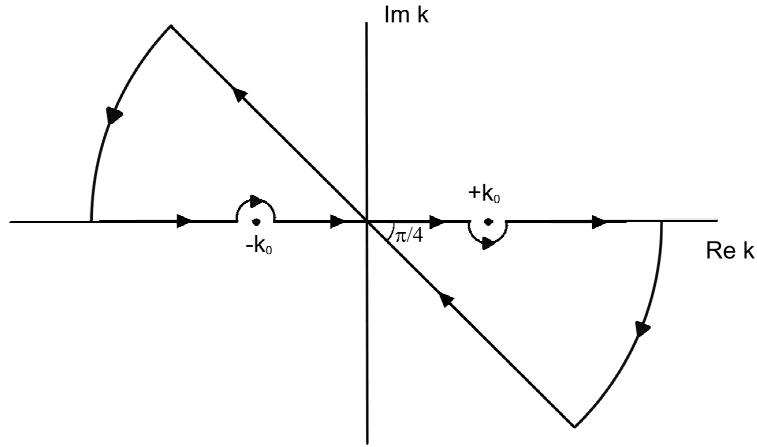}
\caption{\label{fig:loop} The integration contour for evaluating the wave-function (\ref{expansion}) via (\ref{c-answer1}).}
\end{figure}

Next let us turn to the time evolution of the scattering state in the coordinate representation as it is required to define the probability current (\ref{current}). To do so one must plug the amplitude
(\ref{c-answer1}) into expression (\ref{expansion}) (understood in the Cauchy principal value sense). The delta-terms obviously do not present any difficulties while in order to deal with the real axis poles $k=\pm k_0$ we choose the loop in such a way that the exponential $\exp [-i E(k) t/\hbar]$ decays everywhere apart from the real axis and all the poles of the transmission and reflection coefficients are avoided. We have opted for the contour shown in Fig.\ref{fig:loop}. In the limit of large $t$ the integral (\ref{expansion}) over the bisector $\Re[k]=-\Im[k]$ is of the saddle point type with the main contribution coming from the origin and it is straightforward to show that it decays in time as $t^{-3/2}$ (see \ref{sec:appendix-Stationary}). On the other hand the residue contribution in (\ref{c-answer1}) combined
with the delta terms provides a simple asymptotic result for the scattering state:
\begin{eqnarray}
\chi_{k_0}(x,t)&=&\frac{1}{2}\,\left(1+\Re(t_{k_0}-r_{k_0}-r_{k_0}\,t^*_{k_0})\right.)\,e^{i\,k_0\gamma(t)} \psi_{k_0}(x-\gamma(t))\,e^{-\frac{i E(k_0)}{\hbar} t} \nonumber \\
&=& e^{i\,k_0\,\gamma(t)}\, \psi_{k_0}(x-\gamma(t))\,e^{-\frac{i E(k_0)}{\hbar} t},  \quad t \to \infty.
\label{psi-answer}
\end{eqnarray}
But this looks remarkably like the Berry phase result (\ref{Berry}) in the coordinate representation obtained here by taking into account only the singular part of the wave amplitude $c(k,t)$. The
effective Berry phase here is $\phi_{k_0}(t)=k_0 \,\gamma(t)$. The same result one gets by applying the formula for the BP for a continuous eigenstate \cite{ms08} which in our notations reads
$\phi_{k_0} = i\, \gamma(t) \int A(k,k_0;0) dk$ \footnote{Strictly speaking applying the corresponding formula from \cite{ms08} to the kernel (\ref{Akk}) gives rise also to an imaginary part of the phase. The reson for this is that in their derivation the authors of \cite{ms08} assumed that the kernel $A(k,k';0)$ is always diagonal which is clearly not the case for (\ref{Akk}).}. In the same approximation the probability flux built on the wavefunction (\ref{psi-answer}) is just the static probability current of the initial eigenstate
$k_0$:
\begin{equation}
J_{k_0}(x,t)=J_{st}(k_0)=\hbar k_0 |t_{k_0}|^2/(2\pi m), \quad t \to \infty.
\label{current-answer}
\end{equation}
We will call the results (\ref{psi-answer}),(\ref{current-answer}) the Berry phase contribution. Recall that these were obtained in the limit $t\to \infty$ by neglecting the oscillating tails of the amplitude $c(k,t)$.  To see the nature of these fast oscillating tails let us return to Eq.(\ref{ODEs}) and present the
solution as a sum $c(k,t)=c_0(k,t)+\tilde c(k,t)$ where $c_0(k,t)$ is given by Eq.(\ref{c-answer1}) and is the solution of (\ref{ODEs})
if one neglects the time
dependence of the kernel $A$. Then for the non-adiabatic correction $\tilde c(k,t)$ that we assume due to numerical evidence to be a smooth differentiable function one obtains an inhomogeneous equation:
\begin{equation}
\label{tilde-c}
\fl \dot{\tilde c}(k,t)-\dot{\gamma}(t)\int  A(k,k';t) \,\tilde c(k',t) \, dk' = \dot{\gamma}(t)\int  \left[ A(k,k';t)-A(k,k';0) \right]
\,c_0(k',t) \, dk'.
\end{equation}
We can evaluate
the right
hand side exactly using the expressions for $A(k,k',t)$ in Eq.\ (\ref{Akk}) and for $c_0(k,t)$ in Eq.\ (\ref{c-answer1}). As we assume furthermore that $\tilde{c}(k,t)$ possesses no singularities, we
perform on the left hand side the remaining $k'$-integration for large $t$ by closing the integration contour in the complex plane using the same integration loop as in Fig.\ref{fig:loop}. We then obtain the following equations with $k>0$
\begin{eqnarray}\label{complex1}
\fl\dot{\tilde{c}}(k,t)-\dot{\gamma}(t)ik\left(|t_k|^2-|r_k|^2\right) \tilde{c}(k,t)+\dot{\gamma}(t)ik\left(t_kr_k^*-r_kt_k^*\right)\tilde{c}(-k,t)
\\
=\dot{\gamma}(t)\left(e^{i(E(k)-E(k_0))t/\hbar}F(k,k_0,
e^{ik\gamma(t)},e^{ik_0\gamma(t)})+G(k,k_0,
e^{ik\gamma(t)},e^{ik_0\gamma(t)})\right)\nonumber
\end{eqnarray}
and
\begin{eqnarray}\label{complex2}
\fl\dot{\tilde{c}}(-k,t)+\dot{\gamma}(t)ik\left(|t_k|^2-|r_k|^2\right) \tilde{c}(k,t)+\dot{\gamma}(t)ik\left(t_kr_k^*-r_kt_k^*\right)\tilde{c}(k,t)
\\
=\dot{\gamma}(t)\left(e^{i(E(k)-E(k_0))t/\hbar}{\tilde F}(k,k_0,
e^{ik\gamma(t)},e^{ik_0\gamma(t)})+{\tilde G}(k,k_0,
e^{ik\gamma(t)},e^{ik_0\gamma(t)})\right)\nonumber
\end{eqnarray}
with the functions $F(k,k_0,e^{ik\gamma(t)},e^{ik_0\gamma(t)})$, $G(k,k_0,e^{ik\gamma(t)},e^{ik_0\gamma(t)})$ and $\tilde{F}(k,k_0,e^{ik\gamma(t)},e^{ik_0\gamma(t)})$,
$\tilde{G}(k,k_0,e^{ik\gamma(t)},e^{ik_0\gamma(t)})$ obtained by integrating the right hand of Eq.\ (\ref{tilde-c}). We assumed that the function $\tilde{c}(k,t)$ possesses no poles arbitrarily close to the real axis.
Note in this context that the expressions on the right hand side of Eq.\ (\ref{tilde-c}) and Eqs.\ (\ref{complex1},\ref{complex2}) also have no poles in that region as the poles in $c_0(k',t)$ are
canceled by the factor $e^{i(E(k)-E(k'))t/\hbar}-1$ in Eq.\ (\ref{tilde-c}).
The system of differential equations shown above can be diagonalized and the resulting differential equations can be solved in the case
that averages of functions of $\gamma(t)$ are known. The solution can be
represented as time integral with the integrand also containing the inhomogeneities on the right hand side of Eqs.\ (\ref{complex1},
\ref{complex2}).
Due to their time dependence $\propto e^{i(E(k)-E(k_0))t/\hbar}$ also $\tilde{c}(k,t)$ will oscillate in that way in case there are no
significant oscillations from other source, i.e. the noise.

\begin{figure}[h!]
\centering\includegraphics{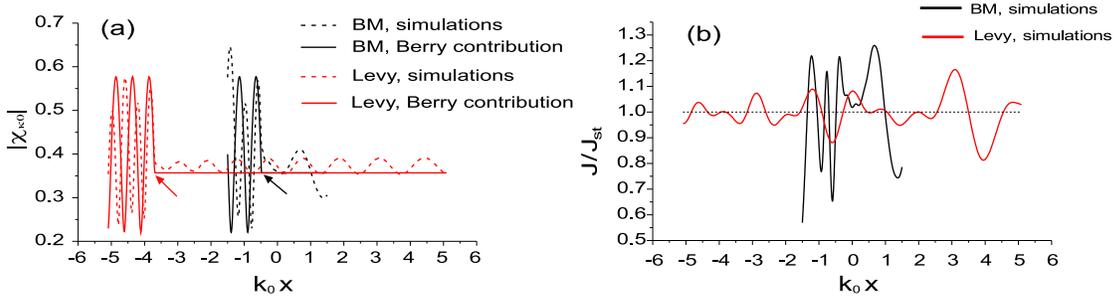}
\caption{\label{fig:psi_current} (a) The simulated scattered state $|\chi_{k_0}(x,t)|$ compared to the BP contribution, (b) the normalized probability current at $t=8$. The deviation of the latter from the static value, (\ref{current-answer}) are due to non-adiabatic effects.}
\end{figure}
All this is illustrated in Fig.\ref{fig:psi_current} where the BP contribution to the scattering state (\ref{psi-answer}) and its probability flux (\ref{current-answer}) are compared with the results of numerical simulations with the same parameters used to obtain Fig.\ref{fig:compare_new}. Only part of the BM solution is shown. In \ref{fig:psi_current}(a) one can immediately discern the final position of the barrier (indicated by arrows) as the shape of the solution is different at the preceding and trailing ends. One notices that the non-adiabatic contribution, $\tilde c$, is important leading in particular to the non-uniform distribution of the probability current in \ref{fig:psi_current}(b) (note however that the fluctuations appear to be smaller for the Levy case).

The latter result should not come as a surprise, after all one should not expect that the Berry phase result obtained adiabatically should hold exactly for a rapidly changing position of the barrier. However
in the next section we shall see that surprisingly when one considers ensemble averaged quantities that are linear in the amplitude $c(k,t)$, in the large time limit Eqs.(\ref{c-answer1}) and (\ref{current-answer}) can be used to determine \textit{the average quantities} while the non-adiabatic correction seems to be always averaged out.

\section{The disorder-averaged quantum propagators}
\label{sec:prop}
As mentioned earlier one important property of the scattering states apart from their being used as the basis for calculating the electrical current (\ref{charge}) is that expressed in the co-moving frame representation $|k,\gamma(t) \rangle$ (\ref{expansion}),(\ref{rotation}) they serve as \textit{quantum propagators}. By these we means the solutions,  $K(k,k';t)$, of Eq.(\ref{ODEs}) given the initial condition $K(k,k';0)=\delta(k-k')$. In the previous section we have already seen that for each realization of the CTRW each of these solution separates into the BP contribution (\ref{c-answer1}) and the non-adiabatic part. In this section we show how one can calculate the \textit{exact average} of the propagator without any a-priori assumptions on the structure of the solution. The methods used here differ slightly for the BM and Levy case but before we consider each case in detail a few remarks are in order. Firstly according to the time-reversal symmetry
mentioned earlier it is only sufficient to consider the averaged propagator $\overline{K}(k,k';t)$ for positive $k'$ as for each realization of $\gamma(t)$ the solution for the negative $k'$ is
obtained from its positive $k'$ counterpart by reversing the direction of motion: $k'\to k$, $\gamma \to -\gamma$. Since we here only consider a symmetric (unbiased) CTRW the average of any function of
$\gamma(t)$ is insensitive to the sign so one gets $\overline{K}(k,-k';t)=\overline{K}(k,k';t)$.

Another important feature shared by all such types of CTRW is that if one is only interested in the average survival amplitude (or Weyl differential) of the initial eigenstate $|k_0 \rangle$
then as we have seen from Fig.\ref{fig:compare_new}(c,d), Eqs.(\ref{Weyl}) are in excellent agreement with the numerics for each realization. Therefore one can average these formulae directly and
arrive at the following amazing  statement: \textit{Regardless of the nature of the symmetric CTRW of a narrow potential barrier the average survival amplitude $\overline{f}(k_0,t)$ of an eigenstate
$|k_0 \rangle$ is just the characteristic function $\Phi(k_0,t)$ of the current position of the scatterer $\gamma(t)$}. This result is more amazing inasmuch as it coincides with the disorder averaged BP
formula (\ref{Berry}) but obtained here in the non-adiabatic regime. In other words: if the system experiences symmetric CTRW the celebrated BP result is recovered \textit{on average}. The
characteristic function of the BM is readily
obtained since $\gamma(t)$ in this case is a zero mean Gaussian variate with the variance $2 D t$. The case of Levy walk is slightly more complicated but one can derive (\cite{KS}, chapter 8) an
asymptote of the time Laplace transform $\tilde \Phi(s,k_0)$ for the case of the Levy-Smirnov distribution ($\alpha = 1/2$) for small $s$ and $v_0 k_0 \tau \ll 1$: $\tilde \Phi(s,k_0) \approx
(s^2+v_0^2 k_0^2)^{-1/2}$. The inverse Laplace transform of the above is known and provides the asymptotic form of the characteristic function as $t \to \infty$. Summarizing the results one has for the
average survival amplitude:
\begin{equation}
\overline f(k_0,t)\equiv  \overline{\lim_{\delta k \to 0} \intop_{k_0-\delta k}^{k_0+\delta k} c(k,t) \,dk}=
\left\{\begin{array}{cc} e^{-D \, k_0^2 \,t}  & \mbox{BM} \\
J_0(k_0 v_0 t), \quad t \gg \tau & \mbox{Levy (-Smirnov)}
\end{array} \right.
\label{survival}
\end{equation}
Note that the average amplitude is purely real owing to the symmetry of the CTRW.

\begin{figure}[h!]
\centering\includegraphics[width=14cm]{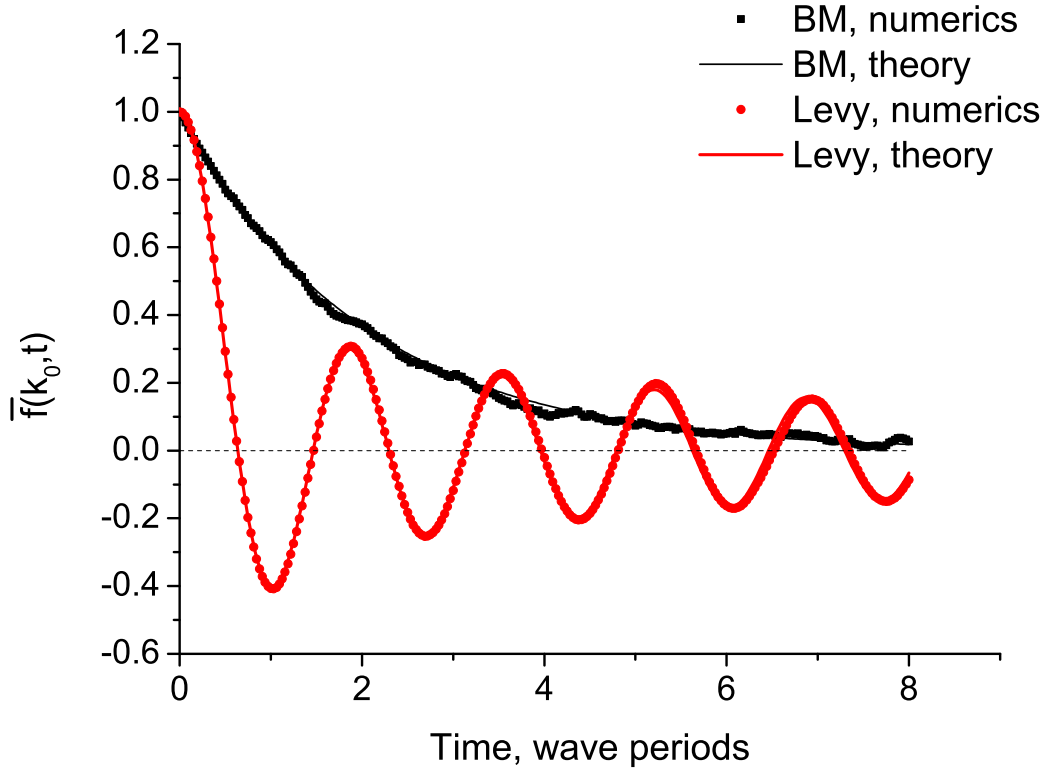}
\caption{\label{fig:survival} The dynamics of the average survival amplitude compared to theoretical prediction. (\ref{survival})}
\end{figure}

To see how well these results are confirmed by the numerics we present a result of averaging over 1000 realizations of the CTRW with the rest of the parameters identical to those of the previous section. The results are given in Fig.\ref{fig:survival}. One can see an excellent agreement between theory and numerics and observe the qualitative difference of the amplitude behaviour: the BM amplitude decays exponentially in time while the Levy walk corresponds to slowly ($\propto 1/t^{1/2}$) decaying oscillations. The imaginary part of the average amplitude (not shown) was observed to be zero up to the third digital place.

So how can one explain that the behaviour at $k_0$ is to a good approximation unaffected by the transitions to other $k$'s? One way to understand it intuitively goes as follows: Starting out at $t=0$ we have $c(k,0)=\delta(k-k_0)$, i.e. the only excited state is $|k_0 \rangle$. From Eq.(\ref{ODEs}) we see that the initial evolution of this state's amplitude is given by the diagonal (singular) part of the kernel, $A(k,k',t)$ which after averaging around $k_0$, Eq.(\ref{survival}), gives exactly the behaviour prescribed by the Berry formula (\ref{Berry}). In the non-adiabatic case the other states are excited very quickly but since the CTRW is symmetric (i.e. unbiased) the \textit{average} total amplitude transfer from the original $k_0$ to other $k$'s and back is almost zero and the main effect again comes from the diagonal ``self-action'' of the amplitude $c(k_0,t)$. For the BM case this can be rigorously justified which is done in Section \ref{sec:Bron}.

\begin{figure}[h!]
\centering\includegraphics[width=14cm]{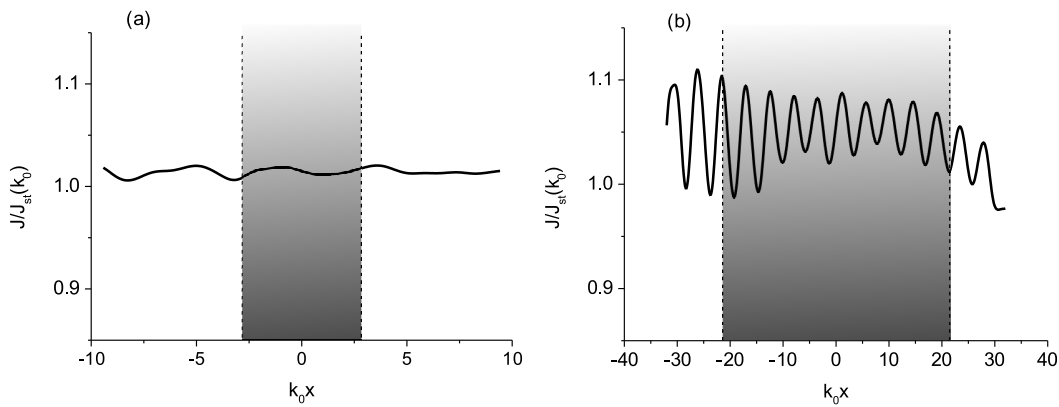}
\caption{\label{fig:curren}  Average probability current normalized to its static value $J_{\rm st}(k_0)$ after
$t = 8$ wave periods, (a)BM case, (b)Levy walk.}
\end{figure}

Besides the averaged survival amplitude also the disorder averaged probability current is of interest. The results for the current are given in Fig.\ \ref{fig:curren}. The averaging was
performed under the same conditions as in the previous sections: 1000 realizations of CTRW were used with the same parameters for each realization as those in the preceding figures.
For both BM and Levy CTRW the average area explored by the potential is shaded in grey.
We can see that for the case of BM the averaged probability current is very close to
its uniform static value which agrees well with the BP result (\ref{current-answer}). On the other hand
for the Levy case the non-adiabatic contributions survive even after averaging leading to
small oscillations of the average current. This is to be expected since
in the Levy case the additional Doppler peaks of the amplitude $c(k;t)$ (see Fig.\ \ref{fig:compare_new}(a)) are not taken into account by Eqs.\ (\ref{c-answer1}) and (\ref{psi-answer}) and
their nonadiabatic contribution leads to the observed spatial oscillations of the average current.

Let us now turn to the averaged propagator and evaluate not only the BP contribution but also the average non-adiabatic diffusion in the $k$-space. We start with the case of the Levy walk.

\subsection{Levy walk}
\label{sec:Levy}

Since in the Levy model the scatterer moves freely between the two jolts for relatively long times it is natural to study the propagator for the barrier moving with constant velocity $v$, $\hat{K}_{v}$. In the coordinate representation one can of course use the Galilei transform of a static propagator $K_{st}(x,x';t)$ of Eq.(\ref{Schrodinger}) for $\gamma \equiv 0$.  The result is
\begin{equation}
\label{Green-x}
K_v(x,x';t)=e^{-i E(\tilde k) t/\hbar} \, K_{st}(x-vt,x';t)\,e^{i\,\tilde k\,(x-x')},
\end{equation}
where $\tilde k = mv/\hbar$ is the wave-vector corresponding to the velocity $v$. On the other hand the explicit expression for the static propagator is known for some important classes of barriers, including delta- and hyperbolic secant ones \cite{gs86,ek88}. For the time evolution of the initial state $|k_0,0\rangle$ in the coordinate representation we then obtain from (\ref{Green-x})
\begin{eqnarray}
\chi_{k_0}(x,t)&=\int_{-\infty}^\infty dx' K_v(x,x';t)\psi_{k_0}(x')\nonumber\\
&=\int_{-\infty}^\infty dx'dke^{-i(E(\tilde{k})+E(k))t/\hbar}e^{i\tilde{k}(x-x')}\psi_k(x-vt)\psi_k^*(x')\psi_{k_0}(x'),
\end{eqnarray}
where we have applied the eigenfunction expansion in the second step. We can perform the $x'$-integration exactly using the explicit expressions for the wave-functions of a delta-scatterer and afterwards
evaluate
the $k$-integrals in the limit of large $t$ using the contour in Fig.\ \ref{fig:loop} to obtain ($k_0,\tilde k>0$ is assumed):
\begin{eqnarray}
\label{psi-v-const}
\fl \chi_{k_0}(x,t)=\left\{\begin{array}{cc}t_{k_0-\tilde{k}}e^{ik_0x}e^{-iE(k_0)t/\hbar}, &x> vt\\
e^{ik_0x}e^{-iE(k_0)t/\hbar}+r_{k_0-\tilde{k}}e^{-i(k_0-2\tilde{k})x}e^{-iE(k_0-2\tilde{k})t/\hbar}, &x<vt\end{array}\right., \quad t \to \infty.
\end{eqnarray}
The wavefunction corresponding to the barrier moving in the opposite direction is obtained via flipping the sign of the corresponding wave-vector $\tilde k$.
Focusing on the transmitted wave or current at $x>0$ we observe that the transmission coefficient $t_{k_0}$ obtained for a static
potential is now replaced by $t_{k_0-\tilde{k}}$ and $t_{k_0+\tilde{k}}$ which inevitably leads to the appearance of electric current.
It is interesting to observe how in the adiabatic limit of small velocity $\tilde k \ll k_F$ one recovers the ``snow-plow'' formula (\ref{Buttiker}) (see \ref{sec:appendix-BTP}).

When one considers a Levy walk however it is more advantageous to work with the expressions for the constant velocity propagator in the co-moving basis (\ref{expansion}),(\ref{rotation}):
\begin{equation}
\label{Green}
K_v(k,k';t)=e^{i\frac{E(\tilde k)\,t}{\hbar}}\,\int dk'' \langle k | e^{i\,\tilde k \hat{x}} | k'' \rangle \, e^{\frac{i}{\hbar}(E(k)-E(k'')) t} \, \langle k'' | e^{-i\,\tilde k \hat{x}} | k' \rangle.
\end{equation}
The easiest way to derive the expression above is to notice that an arbitrary solution of Eq.(\ref{Schrodinger}) with a uniformly moving barrier can be presented as a linear combination of the moving frame solutions $\psi_k(x-vt)\,\exp(i\,\tilde k\,x)$ and use the initial condition $\psi(x,0)=\psi_{k'}(x)$ to find the coefficients.

\begin{figure}[h!]
\centering\includegraphics[width=14cm]{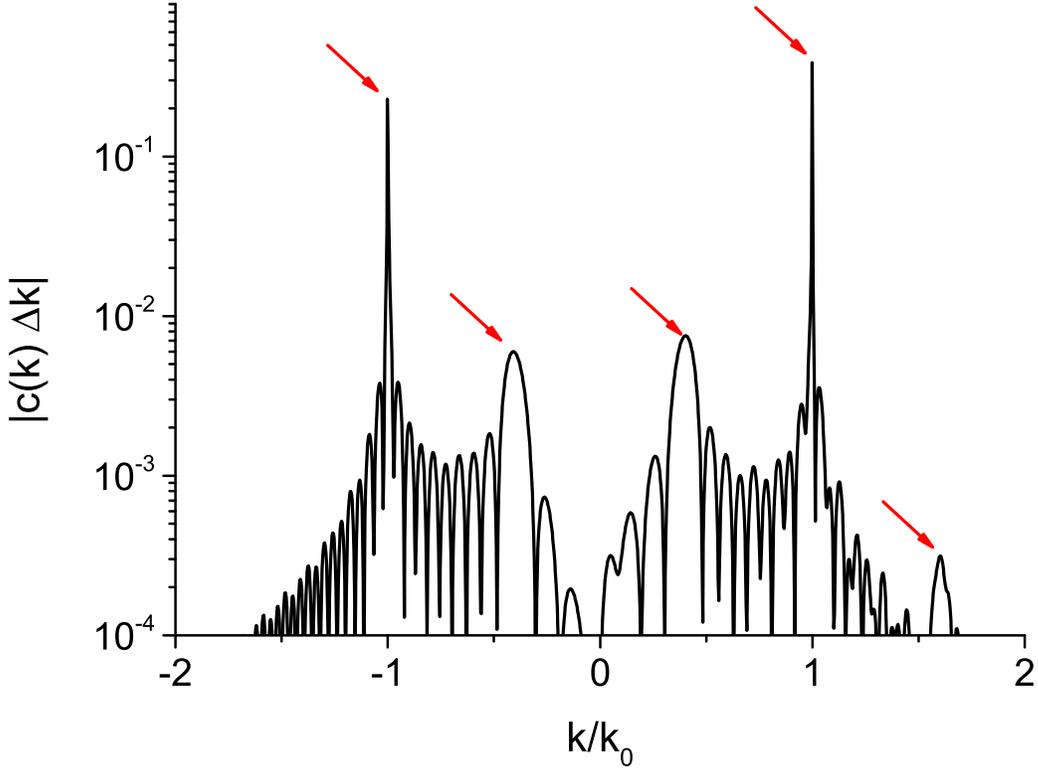}
\caption{\label{fig:c} The amplitude distribution for a uniformly moving scatterer after $t=8$ periods of the moving wave}
\end{figure}

The propagator (\ref{Green}) has some features interesting for the future analysis. First of all matrix elements $ \langle k | \exp(\pm i\,\tilde k \hat{x}) | k' \rangle$ can be calculated
explicitly
for a delta-barrier and they have delta peaks (as well as simple poles) at the Doppler-shifted wave vectors: $|k \pm k'|=\tilde k$. Thus the amplitude $c(k,t)$ initially corresponding to an eigenstate $|k_0\rangle$ ($k_0>0$) will have \textit{six} sharp peaks. Apart from the two main peaks at $k=\pm k_0$ there are four additional ones located symmetrically at the Doppler shifted points $k=\pm(k_0+2\tilde k)$ and $k=\pm |2\tilde k -k_0|$. The two main peaks $k=\pm k_0$ (always dominating in terms of magnitude) experience periodic time oscillations with the frequency $\omega_\pm=(1/\hbar)(E(\tilde k) + E(k_0)-E(k_0\pm \tilde k ))= \pm k_0 v$ which coincides with the BP result (\ref{Berry}). This is also consistent with the long time asymptotic result in the coordinate
domain (\ref{psi-v-const}).
The side peaks however cannot be obtained from a simple adiabatical approach of e.g. \cite{aegs00} and the self-consistent adiabatic treatment must include the  newly created Doppler shifted states as the additional source of pumping for the main mode $k_0=k_F$ (see \cite{jcsf08}). The beating frequencies of the Doppler-shifted peaks are given by the combinations $\omega=(1/\hbar)(E(\tilde k) + E(k_0 \pm 2\tilde k)-E(k_0\pm \tilde k ))$. Note however that unlike Ref.\cite{jcsf08} here these peaks are present in completely non-adiabatic settings. All this is illustrated in a specific example in Fig.\ref{fig:c} in which we considered a right moving scatterer with the same constant velocity as in the Levy walk example in the section (\ref{sec:Berry}) corresponding to $\tilde k=0.3 k_0$. Most of the peaks are clearly visible and indicated by arrows. Note that not all six peaks can be observed at any given moment, e.g. the leftmost peak at $k=-1.6 k_0$ in Fig. \ref{fig:c} is absent.

This picture helps to shed a new light on discrepancies between the $k$-space diffusion for BM and Levy model presented earlier in Fig.\ref{fig:compare_new}(a). Indeed for the Brownian motion the velocity of a particle at
each time moment is sampled from a symmetric Gaussian distribution which has zero mean and a very large variance (inversely proportional to the correlation time, $\tau$). Therefore the effective Doppler wave vector $\tilde k$ experiences large fluctuations around zero at each time moment and therefore no coherent picture of the peaks emerges. The Levy walk on the other hand spends a long time locked in the motion with fixed positive or negative values of the Doppler shift $\pm \tilde k$ which explains why at least some of the peaks from Fig. \ref{fig:c} survive in the stochastic motion in Fig.\ref{fig:compare_new}(a).

Next let us turn to the averaging of the multiple collision propagator, originating from a large number of velocity jumps. For our Levy walk it is the propagator $\hat{K}_v$ (either in coordinate representation (\ref{Green-x}) or in the moving $k$-frame, (\ref{expansion},\ref{Green})) explicitly depending on the time interval and velocity, that serves as the building block for the overall propagator after $n$ discrete random jolts:
\begin{equation}
\label{Green-total}
\hat{K}_n=\prod_{i=0}^{n} \hat{K}_v(\Delta t_i,v_i), \quad \hat{K}_0 \equiv \hat{I}.
\end{equation}
Then for the average propagator at time $t$ one has the expression:
\[
\overline{\hat{K}}(t)=\sum_{n=0}^{\infty} \chi_n(t) \overline{\prod_{i=0}^{n} \hat{K}_v(\Delta t_i,v_i)},
\]
where $\chi_n(t)$ is the probability of having exactly $n$ jolts up to time $t$. The above can be simplified if we recall our assumption that the time intervals and velocity values for each event are independent and identically distributed and the joint velocity-time distribution is \cite{KS}:
\[
P(v,t)=\frac{1}{2}\delta(|v|-v_0)\,p(t),
\]
where $p(t)$ is our heavy-tailed inter-event distribution function.  Then one has:
\begin{equation}
\label{Green-ave}
\overline{\hat{K}}(t)=\sum_{n=0}^{\infty} \chi_n(t) \overline{\hat{K}_v}^n, \quad    \overline{\hat{K}_v} \equiv \left(\frac{\overline{\hat{K}}(\Delta t,v_0)+\overline{\hat{K}}(\Delta t,-v_0)}{2}\right),
\end{equation}
where the averaging is now performed over the random time intervals, $\Delta t$, only. To proceed further we use the standard method in the theory of CTRW, namely work with time Laplace transformed quantities
\cite{KS}. In the Laplace domain for example the probability $\chi_n(t)$ has an especially simple form: $\chi_n(s)=p^n(s)(1-p(s))/s$ \cite{KS}. This allows for writing the average propagator in the Laplace domain in a compact form:
\begin{equation}
\label{Green-ave-Laplace}
\overline{\hat{K}}(s)=\sum_{n=0}^{\infty}  \frac{1-p(s)}{s} p^n(s) \, \overline{\hat{K}_v}^n= \frac{1-p(s)}{s} \, (1-p(s)\overline{\hat{K}_v})^{-1},
\end{equation}
where we have assumed that the geometric series of operators converges. Continuing borrowing from the established methods of the CTRW theory \cite{KS} we note that the large time asymptote of the propagator is
determined by small $s$ behaviour of $\overline{\hat{K}}(s)$. In particular by virtue of the so called Tauberian theorems the small $s$ expansion of the fat tailed PDF
$p(t)\propto \tau^\alpha/t^{1+\alpha}$ is given by $p(s)=1-\tau^\alpha s^\alpha + \ldots$.  Then keeping only the principal terms in $s$ one arrives at $\overline{\hat{K}}(s) \approx \tau^\alpha s^{\alpha-1}(1-\overline{\hat{K}_v})^{-1}$  which back in the time domain corresponds to the asymptote
\begin{equation}
\label{Green-ave-answer}
\overline{\hat{K}}(t)=\left(\frac{\tau}{t}\right)^{\alpha}\frac{1}{\Gamma[1-\alpha]} \,(1-\overline{\hat{K}_v})^{-1}, \quad t \to \infty.
\end{equation}
Finally let us simplify the expression for the average single jump kernel $\overline{\hat{K}_v}$ in the moving basis (\ref{Green}). From the definition of $\overline{\hat{K}_v}$ in Eq.(\ref{Green-ave}) one obtains:
\[
\overline{\hat{K}_v}=\frac{1}{2} \int d k'' \Phi_{k,k'';v_0} \left(\langle k|e^{i \tilde k x}| k''\rangle \langle k''|e^{-i \tilde k x}| k'\rangle+ \langle k|e^{-i \tilde k x}| k''\rangle \langle k''|e^{i \tilde k x}| k'\rangle \right),
\]
where $\Phi_{k,k'';v_0}$ is the characteristic function of the waiting time distribution, i.e. $p(s)$, evaluated at the imaginary argument $s_*=(i/\hbar)(E(\tilde k) +E(k)-E(k''))$. Assuming that we are interested only in the vicinity of the main six peaks (see above), that $\tilde k \sim k_0$ and additionally $\tau E(k_0)/\hbar \ll 1$ one can use the familiar small $s$-expansion of the characteristic function and obtain:
\begin{equation}
\label{Kv-ave}
\fl \langle k| 1-\overline{\hat{K}_v}| k' \rangle = \frac{1}{2} \tau^{\alpha}  \int d k'' s^\alpha_*(k,k'';v_0) \left(\langle k|e^{i \tilde k x}| k''\rangle \langle k''|e^{-i \tilde k x}| k'\rangle+ \langle k|e^{-i \tilde k x}| k''\rangle \langle k''|e^{i \tilde k x}| k'\rangle \right)
\end{equation}
Note that according to Eq.(\ref{Green-ave-answer}) the average propagator decays in time as $t^{-\alpha}$. It is interesting to compare it to the results of Section \ref{sec:Berry} for the
Levy-Smirnov distribution with $\alpha=1/2$ as the average survival amplitude considered there is proportional to the diagonal matrix element $\langle k_0| \overline{\hat{K}}(t) |k_0 \rangle$. One can see both from Eq.(\ref{survival}) and Fig.\ref{fig:survival} that the survival amplitude indeed decays in time as $t^{-1/2}$ as predicted by Eq.(\ref{Green-ave-answer}). What the latter formula fails to predict though are the oscillations of the corresponding Bessel function. To get these oscillations one must keep higher terms in the
$s$-expansion of $(1-p(s)\overline{\hat{K}_v})^{-1}$ in Eq.(\ref{Green-ave-Laplace}). A similar situation in fact occurs when considering the large time limit of the characteristic function of a Levy-Smirnov walk (see Chapter 8 in \cite{KS}). 

\subsection{The Brownian motion}
\label{sec:Bron}

To get the ensemble averaged propagator $\overline{\hat{K}}(k,k';t)$ (in the co-moving representation) for the case of Brownian motion (\ref{Brownian}) we first derive an \textit{exact} evolutionary equation for the average amplitude $c(k,t)$ without assuming anything about the shape of the potential. This is done by averaging Eq.(\ref{ODEs}) directly and applying Furutsu-Novikov formula (corresponding to Stratonovich regularization) \cite{Klyatskin}. The result reads (see \ref{sec:Appendix-averages}):
\begin{equation}
\dot{\bar{c}}(k,t)= D\, \int dk' \, \langle k| \hat A^2(t=0) | k' \rangle e^{i(E(k)-E(k'))t/\hbar} \,\bar{c}(k',t).
\label{c-ave}
\end{equation}
For completeness we have also derived an equation for the binary correlation function $C(k,k';t)=\overline{c^*(k',t)\,c(k,t)}$ (see \ref{sec:Appendix-averages}):
\begin{equation}
\frac{d \hat{C}}{dt}= D\,\left(\hat{A}^2(t)\,\hat{C}+\hat{C} \hat{A}^2(t) -2 \hat{A}(t)\hat{C}\hat{A}(t) \right) = D\, \left[\hat{A}(t),\left[\hat{A}(t),\hat{C}\right]\right],
\label{eq-C}
\end{equation}
where we have used skew-Hermicity of $\hat{A}(t)$. This equation is quite interesting in itself but its detailed analysis lies beyond the scope of this paper.

Let us use (\ref{c-ave}) to calculate the the desired average propagator $\overline{\hat{K}}(k,k_0;t)$ which is the solution of the above equation subject to the initial condition $\bar{c}(k,0)=\delta(k-k_0)$.
In the case of a delta-potential the matrix elements of the operator $\hat{A}^2(t=0)=\partial^2/\partial x^2$ in the basis $|k \rangle$ can be computed analytically
\begin{equation}
\langle k| \hat{A}^2(t=0) | k' \rangle = -k^2\, \delta(k-k')+(\Omega/\pi) t^*_{|k|} \,t_{|k'|}.
\end{equation}
The fastest way to obtain this is to use the stationary Schr\"{o}dinger equation for $\psi_{k'}(x)$ to get its second derivative and then project it onto $\psi_{k}(x)$.

Then from Eq.(\ref{c-ave}) one immediately obtains an \textit{exact} equation for the average coefficients in the moving frame:
\begin{equation}
\fl \dot{\bar{c}}(k,t)=-D\,k^2\,\bar{c}(k,t)+D\, \frac{\Omega}{\pi}\int_{-\infty}^\infty dk' \, \frac{|kk'|e^{i(E(k)-E(k'))t/\hbar}}{(|k|-i\Omega)(|k'|+i\Omega)} \,
\bar{c}(k',t).
\label{c-ave-eq}
\end{equation}
Given the initial condition $\bar{c}(k,0)=\delta(k-k_0)$ the solution can be presented as
\begin{equation}
\label{c-ave-answer}
\fl \overline{\hat{K}}(k,k_0;t) = \bar{c}(k,t)=\delta(k-k_0) \,e^{-D k_0^2 t} +\frac{D \Omega |k|}{\pi(|k|-i\Omega)} \, \intop_0^t e^{-D k^2 (t-t')} \, e^{i E(k) t'/\hbar} \, g(t') \,dt',
\end{equation}
where $g(t)$ is the solution of the Volterra equation of the second kind:
\begin{eqnarray}\label{eqg}
\fl g(t)  \equiv \int \frac{|k'|e^{-i\,E(k')t/\hbar}}{|k'|+i\Omega}  \bar{c}(k',t) dk'=  \frac{k_0 \, e^{-i\,E(k_0)t/\hbar}}{k_0+i\Omega} e^{-Dk_0^2 t}
\nonumber\\
+  \int dk' \, \frac{D\Omega k'^2}{\pi(k'^2+\Omega^2)} \intop_0^t \,dt' e^{-D k'^2 (t-t')-i(E(k')/\hbar)(t-t')} g(t').
\end{eqnarray}
The first term in (\ref{c-ave-answer}) defines the average survival amplitude, $f(k_0,t)$, and one immediately sees that it agrees with
the BP result (\ref{survival}) and the numerics Fig.\ref{fig:survival}. The second term in (\ref{c-ave-answer}) is the averaged non-adiabatic correction
(due to the diffusion in the $k$-space) and in order to define its asymptotic behaviour one needs to know the behaviour of $g(t)$.

The Laplace time-transform is the method of choice and after some simple algebra one gets for the transformed function $g(s)$:
\begin{eqnarray}
\fl g(s)=\frac{k_0}{(k_0+i\Omega)(s+i\tilde{\omega}(k_0))}+\frac{k_0(1-i)D\Omega^2}{(k_0+i\Omega)(s+i\tilde{\omega}(k_0))\left[\sqrt{2s\tilde{\omega}(\Omega)}+(1+i)E(\Omega)/
\hbar\right]},
\end{eqnarray}
where we have introduced the complex frequency $\tilde \omega(k) \equiv E(k)/\hbar -i\, D k^2$.
The time-domain function $g(t)$ can then be obtained by the standard Mellin formula for the inverse Laplace transform. There is a simple pole at $s=-i\tilde{\omega}(k_0)$ in the left half plane.
The contribution from this pole is the oscillating time decaying exponential:
\begin{equation}
g_1(t)=e^{-i\tilde{\omega}(k_0)t}\frac{k_0\Omega\left(\frac{\hbar}{2m}-iD\right)}{\left[i\frac{\hbar\Omega^2}{2m}+\left(\frac{\hbar}{2m}-iD\right)k_0\Omega\right]}.
\end{equation}
A second contribution is obtained from the branch cut that is situated in the complex plane at $s=-\alpha/(2\tilde{\omega}(\Omega))$ with $\alpha\in\mathds{R}^+$. The sum of the two integrals
over the two sides of the brunch cut provides the remaining contribution:
\begin{equation}
\fl g_2(t)=\frac{k_0(1+i)D\Omega^2}{\pi(k_0+i\Omega)\left[\tilde{\omega}(\Omega)\tilde{\omega}(k_0)+\frac{E(\Omega)^2}{\hbar^2}\right]}\int_0^\infty dxe^{-\frac{x^2t}{2\tilde{\omega}
(\Omega)}}\left[\frac{x^2}{x^2-2i\tilde{\omega}(\Omega)
\tilde{\omega}(k_0)}-\frac{x^2}{x^2+2i\frac{E(\Omega)^2}{\hbar^2}}\right].
\end{equation}
It is possible to evaluate this integral in terms of the error function
\begin{eqnarray}
&\fl g_2(t)=\frac{-k_0(1+i)D\Omega^2}{2(k_0+i\Omega)\left(\tilde{\omega}(\Omega)\tilde{\omega}(k_0)+\frac{E^2(\Omega)}{\hbar^2})\right)}\left[\sqrt{-2i\tilde{\omega}(\Omega)\tilde{\omega}(k_0)}
e^{-i\tilde{\omega}(k_0)t}{\rm Erfc}\left(\sqrt{-i t \tilde{\omega}(k_0)}\right)\right.\nonumber\\&\fl\left.+\sqrt{2i}\frac{E(\Omega)}{\hbar}
e^{iE^2(\Omega)t/(\hbar^2\tilde{\omega}(\Omega))}{\rm Erfc}\left(\sqrt{\frac{iE^2(\Omega) t}{\hbar^2\tilde{\omega}(\Omega)}}\right)\right].
\end{eqnarray}
If we plug both $g_1(t)$ and $g_2(t)$ into the expression (\ref{c-ave-answer}) for the average propagator we obtain correspondingly the contribution $\bar{c}_1(k,t)$ from $g_1(t)$
\[
\bar{c}_1(k,t) \propto \frac{|k|}{|k|-i\Omega} \, \frac{e^{-D k_0^2 t} \,e^{(i/\hbar)(E(k)-E(k_0))t}}{\tilde \omega(k)-\tilde \omega(k_0)}
\]
and the contribution $\bar{c}_2(k,t)$ from $g_2(t)$
\begin{eqnarray}
 &\fl\bar{c}_2(k,t)\propto\frac{|k|}{|k|-i\Omega}\left[\sqrt{2i}\frac{E(\Omega)}{\hbar}e^{iE^2(\Omega)t/(\hbar^2\tilde{\omega}(\Omega))+iE(k)t/\hbar}\frac{{\rm Erfc}\left(\sqrt{i\frac{E^2(\Omega)t}{\hbar^2
 \tilde{\omega}(\Omega)}}\right)}{\left(\frac{E^2(\Omega)}{\hbar^2\tilde{\omega}(\Omega)}+\tilde{\omega}(k)\right)}\right.\nonumber\\&\fl\left.+\sqrt{-2i\tilde{\omega}(\Omega)\tilde{\omega}(k_0)}e^{-i\tilde{\omega}(k_0)t
 +iE(k)t/\hbar}\frac{{\rm Erfc}\left(\sqrt{-i\tilde{\omega}(k_0)t}\right)}{\left(\tilde{\omega}(k)-\tilde{\omega}(k_0)\right)}\right].
\end{eqnarray}
The functions $\bar{c}_1(k,t)$ and $\bar{c}_2(k,t)$ are the corrections to the term localized at $k_0$ in Eq.\ (\ref{c-ave-answer}). Both decay with time, $\bar{c}_1(k,t)$ exponentially and $\bar{c}_2(k,t)$ $\propto 1/\sqrt{t}$ for large $t$.

To summarize, Eqs.\ (\ref{Green-ave-answer}) and (\ref{c-ave-answer}) provide the analytical expressions for the disorder-averaged quantum propagator (in the co-moving representation) for the Levy
and BM case, correspondingly. These expressions can be used e.g. to study quantum scattering of narrow wavepackets at a randomly moving scatterer which is much less studied area than the deterministic wavepacket scattering \cite{phs91,bipamdd12}.

\section{The fluctuations of the pumped charge}

\begin{figure}[h!]
\centering\includegraphics[width=14cm]{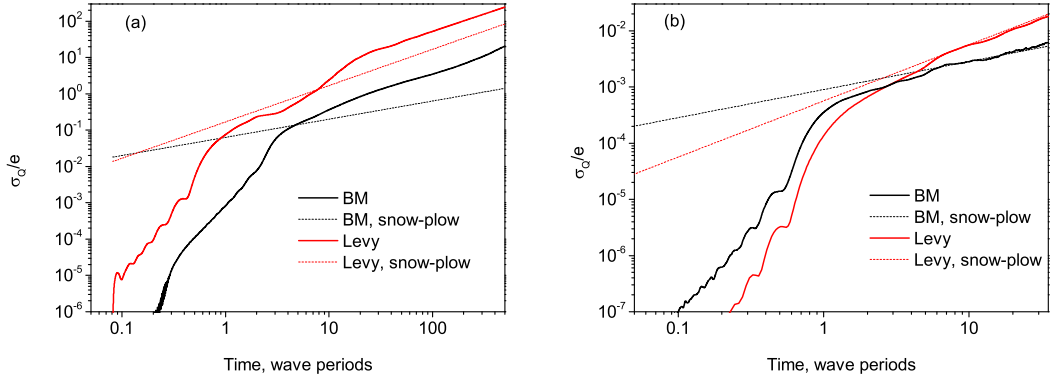}
\caption{\label{fig:charge} The r.m.s of the pumped charge vs time measured in Fermi wave periods $T_F=2 \pi \hbar/E_F$ averaged over $4000$ realizations. (a) Non-adiabatic regime, (b) adiabatic regime.}
\end{figure}

Let us now finally look at the fluctuations of the pumped charge (\ref{charge}).  According to the results of Section \ref{sec:Berry} the BP contribution of each scattering state $\chi_{k_0}$
(\ref{c-answer}) leads to a static probability flux.  On the other hand due to time-reversal symmetry of the static quantum dot potential the absence of bias makes the flow of the electric current impossible without applying an external gate voltage. Therefore it is the non-adiabatic correction to the amplitudes given by Eq. (\ref{tilde-c}) that breaks the time reversal symmetry and introduces a random bias at each realization of the CTRW which in turn leads to the fluctuations of the pumped current and charge. For the unbiased CTRW the average values of both current and charge in (\ref{charge}) should vanish and this is indeed confirmed by the results of the numerical simulations (not shown). Therefore we shall be interested in the time growth of the fluctuations of the pumped charge, namely its RMS value, $\sigma_Q$. This value was sampled at boundary points $\pm L/2$ chosen to be at least three times the standard deviation of the CTRW to avoid crossing effects. Since we have
not noticed significant variation of the results for right and left lead (which is to be expected because of the current conservation) we only provide the results for the right lead $x=+L/2$ which are given in Fig. \ref{fig:charge}.

We see from Fig.\ref{fig:charge}(a) that the time dependence of the standard deviation of the pumped charge shows a few common features for both the BM and Levy case. In particular, one observes that after a short transient time up to $t \sim T_F$ charge fluctuations seem to enter a stable growth phase with the slope larger than predicted by the snow-plow formula (\ref{Buttiker}) i.e. 1/2 for BM and and close to it for the ballistic Levy case. In the latter case however the ballistic spread velocity is still almost an order of magnitude larger than predicted by Eq.(\ref{Buttiker}). We have been able to trace this regime for almost two decades of time variation. Of course the observed deviation from the BTP formula should come as no surprise since the parameters of the simulation used in Fig.\ref{fig:charge}(a) were the same as in the previous chapters and were far from the adiabatic regime. To check what happens in the adiabatic limit we have repeated the 
simulation for a very slow moving quantum dot there the r.m.s. instantaneous wave vector $\tilde k$ was much smaller than $k_F$. The results are given in Fig.\ref{fig:charge}(b). There for the Levy case the effective wave vector was chosen to be $\tilde k =10^{-3} \,k_F$ while for the Brownian motion we defined the instantaneous
velocity via a typical r.m.s. displacement $\tilde v=\sqrt{2D/ \tau}$ where $\tau$ is the effective correlation length imposed by a finite time grid. In our dimensionless units the results of Fig.\ref{fig:charge}(b) are given for $\tau = 4 \times 10^{-3}$, $D=10^{-4}$. As expected, asymptotically the snow-plow effect of Eq.(\ref{Buttiker}) is recovered. Note that the transient period is roughly the same as in the non-adiabatic case of Fig.\ref{fig:charge}(a). In both non-adiabatic and adiabatic cases this transient period corresponds to the time needed for the pump switched at $t=0$ to start working and reflects the inertia of the pumping. In fact in the adiabatic regime this transient behaviour was already numerically observed in Ref.\cite{jcsf08} for slow harmonic oscillations of the barrier strength and position.


As for the large time asymptotes, we observed in the last chapter that the average propagator and the probability current seem to be determined solely by the BP contribution. Despite that, the charge fluctuations (that by virtue of expressions (\ref{expansion}),(\ref{charge}) and (\ref{current}) are determined by the fourth order correlation functions of the reduced amplitudes $\tilde c_\pm(k,t)=c(k,t)\exp(\pm \, i k \gamma(t))$) are non-negligible and yield the charge fluctuations that grow faster than expected from the snow-plow formula (\ref{Buttiker}). The exact theory of the phenomenon will require developing a kinetic description of the sums (integrals) of the fourth order correlation functions built on the $\tilde c_\pm (k,t)$ which will present some theoretical challenges due to the non-Gaussianity of their statistics. This however is beyond the scope of the present paper.


\section{Conclusions and outlook}
\label{sec:Conclusion}
In this paper we have studied the tunneling of a quantum particle through a randomly moving quantum dot and have shown that in the large time limit the disorder-averaged quantum state of a system behaves largely as if one simply
applies the corresponding random geometric phase as is the case for the adiabatic motion of the scatterer. We have also studied the growth of the variance of the the pumped quantum charge (\ref{charge}) and found significant deviations from the ``snow-plow'' contribution (\ref{Buttiker}) derived from an adiabatic BTP formula. We note that most of the results of this paper should apply to scatterers of an arbitrary (but well localized) shape provided
that the typical spatial extent of the latter, $\Delta$, is such that the inequality $k_F \Delta  \ll 1$ holds. While we only report here the results for the two random models Brownian motion and Levy walk we note that the results of the Chapter \ref{sec:Berry} (the separation of the BP contribution and the fast non-adiabatic fluctuations) are valid for an arbitrary random motion of the scatterer. Therefore we still expect that the average survival amplitude is asymptotically given by the characteristic function of the random walk, although the specific shape of it is of course process-dependent (see Eq.(\ref{survival}) and Fig.\ref{fig:survival}).

Concerning possible experimental realizations and especially the observation of the nonadiabatic Berry phase it is interesting to consider the analogy with optics, more precisely the propagation of an
optical beam in a narrow randomly curved waveguide \cite{l09} where an equation of the form (\ref{Schrodinger}) is valid with time variable $t$ replaced by the paraxial propagation distance $z$.  Another interesting possibility arises from the unitary mapping of the considered problem of a randomly moving quantum dot to that of scattering at a spatially fixed barrier exposed to an external temporally fluctuating potential via a Kramers-Henneberger type transformation \cite{hdr00}. As for our results for quantum pumping of electrical charge, note that we get for the BM and Levy model for the values of parameters chosen rather high values of the charge fluctuations up to a few dozens of $e$. We think that such fluctuations should be observable experimentally in quantum dot experiments.

Our work can be extended in various directions on the theoretical side: It would be highly desirable to quantify better the corrections determined by Eq.\ (\ref{tilde-c}) to the Berry phase result for the wavefunction and by doing so understand better the time dependence of the pumped charge fluctuations exceeding the predictions of the adiabatic theory. Furthermore testing our results for extended scattering potentials and considering the pumping of spin instead of charge would be interesting.

We are grateful to Uzy Smilansky for initially drawing our attention to the problem of random tunneling and to Nick Korabel for useful comments on the Levy walk model. SD was supported by the Marie
Curie Fellowship (project ``INDIGO'') and wishes to acknowledge kind hospitality of the Faculty of Physics of the University of Duisburg-Essen. DW acknowledges the Minerva foundation for a fellowship
during the time the project was initiated.

\appendix

\section{The saddle point contribution to the wavefunction}
\label{sec:appendix-Stationary}

The result in Eq.\ (\ref{psi-answer}) is calculated from $c(k,t)$ by performing the $k$-integration along the contour shown in Fig.\ \ref{fig:loop} under the assumption that corrections collected along
the path away from the real axis are zero. We want to determine here the contribution of these corrections for large times $t$. In this case we consider a saddle point approximation around $k=0$.

First one notices that the contributions from the two arches can be made exponentially vanishing by increasing the radius to infinity and one is left only with the contribution from the bisector:
\begin{eqnarray}
&\fl \delta\chi_{k_0}(x,t)=\int_0^\infty dk\,\frac{1-i}{\sqrt{2}}c(k(1-i)/\sqrt{2},t)t_{k(1-i)/\sqrt{2}}{\rm e}^{ik(1-i)(x-\gamma(t))/\sqrt{2}} {\rm e}^{-\hbar k^2t/(2m)}
\\&\fl+\int_{-\infty}^0 dk\,\frac{(1-i)}{\sqrt{2}}c(k(1-i)/\sqrt{2},t)\left({\rm e}^{ik(1-i)(x-\gamma(t))/\sqrt{2}}+r_{(1-i)k/\sqrt{2}}^*{\rm e}^
{-ik(1-i)(x-\gamma(t))/\sqrt{2}}\right) {\rm e}^{-\frac{\hbar k^2t}{2m}}\nonumber
\end{eqnarray}
with $c(k,t)$ given by the last three lines in Eq.\ (\ref{c-answer1}).
We now expand the prefactor of ${\rm e}^{-\hbar k^2t/(2m)}$ around the stationary point $k=0$.
The first two orders in $k$ yield zero, up to second order we obtain for $k>0$
\begin{eqnarray*}
\label{exansion}
 &\fl (1-i)c((1-i)k/\sqrt{2},t)t_{(1-i)k/\sqrt{2}}e^{ik(1-i)(x-\gamma(t))/\sqrt{2}}/\sqrt{2}\approx \nonumber\\&\fl\frac{k^2}{2}\frac{(1+i)}{\sqrt{2}}\frac{\left[k_0\left(e^{ik_0\gamma(t)}-1\right)-2\Omega k_0
 \gamma(t)\theta(\gamma(t))-2\Omega\sin(k_0\gamma(t))\theta(-\gamma(t))\right]}{\pi k_0^2\Omega(k_0+i\Omega)}
\end{eqnarray*}
 and for $k<0$
 \begin{eqnarray*}
\label{exansion1}
 &\fl (1-i)c((1-i)k/\sqrt{2},t)\left({\rm e}^{ik(1-i)(x-\gamma(t))/\sqrt{2}}+r_{(1-i)k/\sqrt{2}}^*{\rm e}^{-ik(1-i)(x-\gamma(t))/\sqrt{2}}\right)/\sqrt{2}\approx
 \nonumber\\&\fl\frac{k^2}{2}\frac{(1+i)}{\sqrt{2}}\frac{2\left[x-\gamma(t)+1/(2\Omega)\right]}{\pi k_0^2
 (k_0+i\Omega)}\times\left[k_0\left(e^{ik_0\gamma(t)}-1\right)+2\Omega k_0
 \gamma(t)\theta(-\gamma(t))\right.\nonumber\\&\fl\left.-2\Omega\sin(k_0\gamma(t))\theta(-\gamma(t))\right].
\end{eqnarray*}
Using the identity
\begin{equation*}
 \int_0^\infty dk\,k^2e^{-\hbar k^2t/(2m)}=\frac{\sqrt{\pi}}{4}\left(\frac{2m}{\hbar t}\right)^{3/2}
\end{equation*}
the final result for $\delta\chi_{k_0}(x,t)$ reads
\begin{eqnarray}
\label{exan}
&\fl \delta\chi_{k_0}(x,t)\approx \frac{\sqrt{\pi}}{4}\frac{(1+i)}{\sqrt{2}}\left(\frac{2m}{\hbar t}\right)^{3/2}\frac{1}{\pi k_0^2(k_0+i\Omega)}\left\{\left[k_0\left(e^{ik_0\gamma(t)}-1\right)-2\Omega\sin(k_0\gamma(t))\theta(-\gamma(t))
\right]\right.\nonumber\\&\fl\left.\left[x-\gamma(t)+1/\Omega\right]- k_0\gamma(t)\left[\theta(\gamma(t))-\theta(-\gamma(t))\right]+2\Omega k_0\gamma(t)
\theta(-\gamma(t))(x-\gamma(t))\right\}
\end{eqnarray}
The expression above for $\delta\chi_{k_0}(x,t)$ is the leading order correction to the wavefunction in $t$ resulting from the $k$-integration for obtaining $\chi_{k_0}(x,t)$ from $c(k,t)$. Higher orders in $1/t$ are
obtained by expanding $c(k,t)$ to higher orders in $k$. 

Concerning the probability flux (\ref{current}), we can evaluate the leading order current correction in $t$
\begin{equation}
\Delta J_{k_0}= \frac{\hbar}{m}\Im\left[\chi_{k_0}^*(x,t)\frac{\partial}{\partial x}\delta\chi_{k_0}(x,t)+\delta\chi_{k_0}^*(x,t)\frac{\partial}{\partial x}\chi_{k_0}(x,t)\right]
\end{equation}
with $\chi_{k_0}(x,t)$ denoting the wavefunction obtained within Berry phase approximation in (\ref{psi-answer}). These contributions are oscillatory in dependence of $x$ and maximally
increase with $x$ like $(x\sin k_0x)$.

\section{The derivation of the ``snow plow'' formula for a slowly moving scatterer}

\label{sec:appendix-BTP}

Let us show that for a barrier moving with constant small velocity the asymptotic result (\ref{psi-v-const}) reproduces correctly the adiabatic BTP formula (in particular the ``snow-plow'' result by Avron et. al. \cite{aegs00}).
According to Eq.(\ref{charge}) one needs to integrate the probability flux of ``left'' and ``right'' scattering states under the 1D Fermi surface to get the electric current at a given point. For a ``left'' scattering state with $k_0>0$ we have already obtained an asymptotic formula (\ref{psi-v-const}). In order to find the ``right'' state we again use time reversal transformation i.e. the fact that if $\psi(x,t;v)$ is a
solution of the time-dependent Schr\"odinger equation (\ref{Schrodinger}) with an even potential then $\psi(-x,t;-v)$ is also a solution. At $t=0$ and negative $k_0$ one has
$\chi_{k_0}(x,0)=\chi_{|k_0|}(-x)$. Therefore the ``right'' scattering state can be obtained via its ``left'' counterpart (\ref{psi-v-const}) by replacing $k_0\to |k_0|$, $x \to -x$, $\tilde k \to
-\tilde k$:
\begin{eqnarray}
\label{chiright}
\fl \chi_{k_0}(x,t)=\frac{1}{\sqrt{2\pi}}\left\{\begin{array}{cc}t_{|k_0|+\tilde{k}}e^{-i|k_0| x}e^{-iE(k_0)t/\hbar}, & x \to -\infty \\
e^{-i|k_0| x}e^{-iE(k_0)t/\hbar}+r_{|k_0|+\tilde{k}}e^{i(|k_0|+2\tilde{k})x}e^{-iE(|k_0|+2\tilde{k})t/\hbar}, & x \to +\infty \end{array}\right.
\end{eqnarray}
for $t\to \infty$ and $k_0<0$.
Assuming that the pumped charge is sampled at a distant right lead $x=L \to + \infty$ and plugging (\ref{psi-v-const}) and (\ref{chiright}) into the equation for the charge (\ref{charge}) one gets:
\[
\fl \frac{d Q}{dt} = - \frac{e \hbar}{2\pi m} \intop_{0}^{k_F} \Im \left[i k_0 |t_{k_0-\tilde k}|^2 +i(-k_0+(k_0+2\tilde k)|r_{k_0+\tilde k}|^2+ \right.
\]
\[
\left. 2\tilde k r_{k_0+\tilde k} e^{2 i(k_0+\tilde k)L} \,e^{i(E(k_0)-E(k_0+2\tilde k))t/\hbar}) \right] dk_0.
\]
When we consider the adiabatic limit where $\tilde k \ll k_F$, one can expand the integrand leaving only the terms linear in $\tilde k$. Then one has (recall that $|t_k|^2+|r_k|^2=1$ ):
\begin{eqnarray}
\fl \frac{d Q}{dt} & \approx &  -\frac{e \hbar}{2\pi m} \intop_{0}^{k_F} \left[ k_0 |t_{k_0}|^2-\frac{\partial |t_{k_0}|^2}{\partial k} k_0\tilde k +k_0(-1+|r_{k_0}|^2) +2\tilde k |r_{k_0}|^2 \right. \nonumber \\
\fl &+& \left. \frac{\partial |r_{k_0}|^2}{\partial k} k_0 \tilde k+2\tilde k \Re \left(r_{k_0} e^{2 i k_0 L} \right) \right] \, d k_0 =-\frac{e \hbar \tilde k}{\pi m} \left[ k_F |r_{k_F}|^2 + \Re \intop_0^{k_F} r_{k_0} e^{2 i k_0 L} dk_0 \right].
\label{charge1}
\end{eqnarray}
The integral term in the last equation decays as $1/L$ and thus becomes negligible for $L$ larger than a few multiples of the Fermi wavelength $2\pi/k_F$.

Finally, after an infinitesimally small time interval $\Delta t$ the scatterer moves the distance $\Delta \gamma = (\hbar \tilde k/m) dt$ and from Eq.(\ref{charge1}) one gets exactly the snow-plow formula (\ref{Buttiker}).

\section{The average amplitude and the correlation function for the Brownian motion}
\label{sec:Appendix-averages}

We start from the derivation of Eq.(\ref{c-ave}). From Eq.(\ref{ODEs}) it follows that
\begin{equation}
\dot{c}(k,t)=\xi(t) \,\int dk' A(k,k';t)\,c(k',t).
\label{ODEs-alpha}
\end{equation}
To average this equation over noise one can use the celebrated Furutsu-Novikov (FN) formula \cite{Klyatskin} which for the white Gaussian noise $\xi(t)$ has the form:
\begin{equation}
\overline{\xi(t) \, R[\xi]}=2D\, \overline{\frac{\delta R}{\delta \xi(t)}},
\label{FN}
\end{equation}
where $R[\xi]$ is an arbitrary functional of the noise. To apply this formula we need to calculate the functional derivative of the solution with respect to noise:
\[
\fl
 \begin{array}{ll} \frac{\delta c(k,t)}{\delta \xi(t)}&= \lim_{t'\to t} \frac{\delta c(k,t)}{\delta \xi(t')}= \lim_{t'\to t} \theta[t-t'] \,\left[\int dk' A(k,k';t')\,c(k',t')\right]\\
 &= \frac{1}{2}\, \int A(k,k';t)\, c(k',t) dk',
 \end{array}
\]
where we have used causality and defined the value of the Heaviside function $\theta[0]=1/2$ which corresponds to Stratonovich convention. Applying the above to the r.h.s. of (\ref{ODEs-alpha}) Eq.(\ref{c-ave}) follows.

To get an equation for the correlation function we can follow a similar procedure as above. First we multiply the l.h.s of Eq.(\ref{ODEs}) by $c^*(k',t)$ add the hermitian conjugate and average over noise. We obtain:
\[
\dot{C}(k,k';t)=\int dk'' A(k,k'';t) \, \overline{\xi(t)c(k'',t)c^*(k',t)} +h.c..
\]
Using FN formula (\ref{FN}) and the variational derivatives $\delta c(k,t)/\delta \xi(t)$  obtained above one arrives at Eq.(\ref{eq-C}).

\end{document}